# The Nu Class of Low-Degree-Truncated, Rational, Generalized Functions. Ia. MINOS for IMSPE Evaluation and Optimal-IMSPE-Design Search

Selden Crary[1], Tatiana Nizhegorodova[2], and Michael Saunders[3*]

1. Palo Alto, CA, USA

2. Livermore, CA, USA

3. Department of Management Science and Engineering
Stanford University, Stanford, CA, USA

**Abstract**

We provide compact algebraic expressions that replace the lengthy symbolic-algebra-generated integrals $I_6$ and $I_8$ in Part I of this series of papers [1]. The "MRSE" entries of Part I, Table 4.3 are thus updated to algebraic expressions. We use MINOS [2-4] to tabulate several IMSPE-optimal designs with one factor and one or two design points, for the exponential-, two of the Matérn-, and the Gaussian-correlation functions, i.e. for the class of problems considered in Part I. The tabulated results can be used as standards for optimal-design-software developers and users.

## 1. Introduction

The present paper is Part Ia of the Roman-numeralled series Parts I through V, with sub-parts, reporting research into what we have dubbed "Nu-class generalized functions," [1], of which the IMSPE objective in the field of statistical design of computer experiments was the original example.

Part I, Appendices I and K of this series of papers included lengthy, verbatim, closed-form, symbolic-manipulation-software-generated, algebraic expressions of one-dimensional integrals, named $I_6$ and $I_8$, of the following products of Matérn-covariance functions, with respective class parameters $\nu = 3/2$ and $5/2$ [1]:

$$\nu = 3/2:\ I_6 \equiv \tfrac{1}{2}\int_{-1}^{1}\left[1+\sqrt{3\theta(a-x)^2}\right]\left[1+\sqrt{3\theta(b-x)^2}\right]e^{-\sqrt{3\theta}(|a-x|+|b-x|)}\,dx.$$

$$\nu = 5/2:\ I_8 \equiv \tfrac{1}{2}\int_{-1}^{1}\left[1+\sqrt{5\theta(a-x)^2}+\tfrac{5\theta(a-x)^2}{3}\right]\left[1+\sqrt{5\theta(b-x)^2}+\tfrac{5\theta(b-x)^2}{3}\right]e^{-\sqrt{5\theta}(|a-x|+|b-x|)}\,dx.$$

The expressions for these integrals in Part I, albeit lengthy, obviated the need for computationally expensive Monte-Carlo approximations that previously had been common practice [5], when the integrals were required, as in the computation of an IMSPE-optimal design based on one of these Matérn-covariance functions.

In the present paper, we reduce these integrals to short algebraic expressions useful for software development involving the IMSPE. These expressions can be considered replacements for the placeholder "MRSE" (machine-readable symbolic expression) in Part I, Table 4.3.

---

We also demonstrate use of the general-purpose optimizer for constrained optimization, *MINOS* [2-4], coupled with our own software for evaluation of the IMSPE. An appendix provides numerical examples of IMSPE-optimal designs found using double- and quadruple-precision-arithmetic systems, *dMINOS+IMSPE* and *qMINOS+IMSPE*, respectively. High-precision examples provide software developers and users means to check the validity of their IMSPE-optimality code.

## 2. Outline

## 3. Symmetry operators $\mathcal{S}_w$ and $\mathcal{T}_{a;b}^{(+)}$

We reintroduce the algebraic symmetry operator $\mathcal{S}_w$ defined in Part I, Sub-appendix R.5 of this series of papers [1], and we define a new operator, $\mathcal{T}_{a;b}^{(+)}$.

$\mathcal{S}_w f(w, x) \equiv f(-w, x)$ changes the sign on all algebraic quantities denoted by $w$, to its right. Example: $\mathcal{S}_w(a + bw + cw^2 + dx + ex^2) = a - bw + cw^2 + dx + ex^2$, where $x$ has no dependence on $w$.

$\mathcal{T}_{a;b}^{(+)} \equiv \mathcal{I} + \mathcal{S}_a \mathcal{S}_b$, where $\mathcal{I}$ is the identity operator. Example: $\mathcal{T}_{a;b}^{(+)}\left(1 + \frac{a+b}{2}\right) = 2$.

## 4. Integral $I_6$

The right-hand side of Part I, Eq. I.1 [1] was the symbolic-manipulation-software (SMS) evaluation of the following integral involving the Matérn-covariance function with class parameter 3/2, under the assumption $-1 \leq a \leq b \leq 1$:

$$I_6 \equiv \frac{1}{2}\int_{-1}^{1} \left[1 + \sqrt{3\theta(a-x)^2}\right]\left[1 + \sqrt{3\theta(b-x)^2}\right] e^{-\sqrt{3\theta(a-x)^2} - \sqrt{3\theta(b-x)^2}} \, dx \,.$$

The result was the following, which is repeated for the reader's convenience:

$$I6 := -\frac{1}{24} \frac{1}{e^{\sqrt{3}\sqrt{\theta}\,a} e^{\sqrt{3}\sqrt{\theta}\,b} \left(e^{\sqrt{3}\sqrt{\theta}}\right)^2 \sqrt{\theta}} \Big(5\sqrt{3} + 18\sqrt{\theta} + 6\sqrt{3}\,b\,\theta$$

$$+ 24\sqrt{3}\left(e^{\sqrt{3}\sqrt{\theta}\,a}\right)^2 \theta\,a\,b \left(e^{\sqrt{3}\sqrt{\theta}}\right)^2$$

$$+ 6\sqrt{3}\left(e^{\sqrt{3}\sqrt{\theta}\,a}\right)^2 \left(e^{\sqrt{3}\sqrt{\theta}\,b}\right)^2 a\,b\,\theta + 6\sqrt{3}\,a\,\theta$$

$$- 9\left(e^{\sqrt{3}\sqrt{\theta}\,a}\right)^2 \left(e^{\sqrt{3}\sqrt{\theta}\,b}\right)^2 a\sqrt{\theta} + 6\left(e^{\sqrt{3}\sqrt{\theta}\,a}\right)^2 \left(e^{\sqrt{3}\sqrt{\theta}}\right)^2 a^3 \theta^{3/2}$$

$$- 6\left(e^{\sqrt{3}\sqrt{\theta}\,a}\right)^2 \left(e^{\sqrt{3}\sqrt{\theta}}\right)^2 b^3 \theta^{3/2} + 30\,a\left(e^{\sqrt{3}\sqrt{\theta}\,a}\right)^2 \left(e^{\sqrt{3}\sqrt{\theta}}\right)^2 \sqrt{\theta}$$

$$- 30\left(e^{\sqrt{3}\sqrt{\theta}\,a}\right)^2 \left(e^{\sqrt{3}\sqrt{\theta}}\right)^2 b\sqrt{\theta} + 6\sqrt{3}\left(e^{\sqrt{3}\sqrt{\theta}\,a}\right)^2 \left(e^{\sqrt{3}\sqrt{\theta}\,b}\right)^2 \theta$$

$$- 9\left(e^{\sqrt{3}\sqrt{\theta}\,a}\right)^2 \left(e^{\sqrt{3}\sqrt{\theta}\,b}\right)^2 b\sqrt{\theta} + 6\sqrt{3}\,a\,b\,\theta + 6\sqrt{3}\,\theta$$

$$- 18\left(e^{\sqrt{3}\sqrt{\theta}\,a}\right)^2 \left(e^{\sqrt{3}\sqrt{\theta}}\right)^2 a^2\,b\,\theta^{3/2}$$

$$+ 18\left(e^{\sqrt{3}\sqrt{\theta}\,a}\right)^2 \left(e^{\sqrt{3}\sqrt{\theta}}\right)^2 a\,b^2\,\theta^{3/2}$$

$$- 12\sqrt{3}\left(e^{\sqrt{3}\sqrt{\theta}\,a}\right)^2 \theta\,a^2 \left(e^{\sqrt{3}\sqrt{\theta}}\right)^2$$

$$- 12\sqrt{3}\left(e^{\sqrt{3}\sqrt{\theta}\,a}\right)^2 \theta\,b^2 \left(e^{\sqrt{3}\sqrt{\theta}}\right)^2$$

$$- 6\sqrt{3}\left(e^{\sqrt{3}\sqrt{\theta}\,a}\right)^2 \left(e^{\sqrt{3}\sqrt{\theta}\,b}\right)^2 a\,\theta - 6\sqrt{3}\left(e^{\sqrt{3}\sqrt{\theta}\,a}\right)^2 \left(e^{\sqrt{3}\sqrt{\theta}\,b}\right)^2 b\,\theta$$

$$+ 18\left(e^{\sqrt{3}\sqrt{\theta}\,a}\right)^2 \left(e^{\sqrt{3}\sqrt{\theta}\,b}\right)^2 \sqrt{\theta} + 5\sqrt{3}\left(e^{\sqrt{3}\sqrt{\theta}\,a}\right)^2 \left(e^{\sqrt{3}\sqrt{\theta}\,b}\right)^2$$

$$- 10\sqrt{3}\left(e^{\sqrt{3}\sqrt{\theta}\,a}\right)^2 \left(e^{\sqrt{3}\sqrt{\theta}}\right)^2 + 9\,a\sqrt{\theta} + 9\,b\sqrt{\theta}\Big)$$

After typesetting and steps of reduction, this integral becomes

$$I_6 = -\frac{1}{24\sqrt{\theta}e^{2\sqrt{3\theta}\left(1+\frac{a+b}{2}\right)}} \left\{ \begin{bmatrix} -10\sqrt{3} \\ -30(b-a)\sqrt{\theta} \\ -12\sqrt{3}(a^2-2ab+b^2)\sqrt{\theta}^2 \\ -6(b^3-a^3+3a^2b-3ab^2)\sqrt{\theta}^3 \end{bmatrix} e^{2\sqrt{3\theta}(1+a)} \right. \\ + \begin{bmatrix} 5\sqrt{3} \\ +9(2+a+b)\sqrt{\theta} \\ +6\sqrt{3}(1+a+b+ab)\sqrt{\theta}^2 \end{bmatrix} \\ \left. + \begin{bmatrix} 5\sqrt{3} \\ +9(2-a-b)\sqrt{\theta} \\ +6\sqrt{3}(1-a-b+ab)\sqrt{\theta}^2 \end{bmatrix} e^{2\sqrt{3\theta}(a+b)} \right\}. \quad (4.1)$$

We can merge the exponents associated with the top square bracket, via

$$\frac{e^{2\sqrt{3\theta}(1+a)}}{e^{2\sqrt{3\theta}\left(1+\frac{a+b}{2}\right)}} = e^{-2\sqrt{3\theta}\left(\frac{b-a}{2}\right)}. \tag{4.2}$$

The last two square-bracketed terms of Eq. 4.1 are of the following form that uses the symmetry operators of Sec. 3:

$$\frac{1}{e^{2\sqrt{3\theta}\left(1+\frac{a+b}{2}\right)}}\left[f(a,b) + e^{2\sqrt{3\theta}(a+b)}\mathcal{S}_a\mathcal{S}_b f(a,b)\right]$$

$$= f(a,b)e^{-2\sqrt{3\theta}\left(1+\frac{a+b}{2}\right)} + e^{-2\sqrt{3\theta}\left(1+\frac{a+b}{2}\right)}e^{2\sqrt{3\theta}(a+b)}\mathcal{S}_a\mathcal{S}_b f(a,b)$$

$$= f(a,b)e^{-2\sqrt{3\theta}\left(1+\frac{a+b}{2}\right)} + e^{-2\sqrt{3\theta}\left(1-\frac{a+b}{2}\right)}\mathcal{S}_a\mathcal{S}_b f(a,b)$$

$$= f(a,b)e^{-2\sqrt{3\theta}\left(1+\frac{a+b}{2}\right)} + \mathcal{S}_a\mathcal{S}_b f(a,b)e^{-2\sqrt{3\theta}\left(1+\frac{a+b}{2}\right)}$$

$$= (1 + \mathcal{S}_a\mathcal{S}_b)f(a,b)e^{-2\sqrt{3\theta}\left(1+\frac{a+b}{2}\right)}. \tag{4.3}$$

Combining Eqs. (4.1)-(4.3) and noting $b \geq a$ gives

$$I_6 = -\frac{1}{24\sqrt{\theta}}\left\{\begin{bmatrix} -10\sqrt{3} \\ -30|b-a|\sqrt{\theta} \\ -12\sqrt{3}|b-a|^2\sqrt{\theta}^2 \\ -6|b-a|^{23}\sqrt{\theta}^3 \end{bmatrix}e^{-2\sqrt{3\theta}\left|\frac{b-a}{2}\right|} \\ +(1+\mathcal{S}_a\mathcal{S}_b)\begin{bmatrix} 5\sqrt{3} \\ +9(2+a+b)\sqrt{\theta} \\ +6\sqrt{3}(1+a+b+ab)\sqrt{\theta}^2 \end{bmatrix}e^{-2\sqrt{3\theta}\left(1+\frac{a+b}{2}\right)}\right\}.$$

Multiplying numerator and denominator by $-\sqrt{3}$, factoring, and using $\mathcal{T}_{a;b}^{(+)}$ from Sec. 3 gives

$$I_6 = \frac{1}{24\sqrt{3\theta}}\left\{2\begin{bmatrix} 15 \\ +15|b-a|\sqrt{3\theta} \\ +6|b-a|^2\sqrt{3\theta}^2 \\ +|b-a|^3\sqrt{3\theta}^3 \end{bmatrix}e^{-2\sqrt{3\theta}\left(\frac{b-a}{2}\right)} \\ -3\mathcal{T}_{a;b}^{(+)}\begin{bmatrix} 5 \\ +3(2+a+b)\sqrt{3\theta} \\ +2(1+a+b+ab)\sqrt{3\theta}^2 \end{bmatrix}e^{-2\sqrt{3\theta}\left(1+\frac{a+b}{2}\right)}\right\}. \tag{4.4}$$

Eq. 4.4 is invariant under interchange of $a$ and $b$ and thus it is valid not only under the original assumption $b \geq a$ but also for $b < a$. Thus, Eq. 4.4 is valid for arbitrary $a$ and $b$ on the interval $-1 \leq a, b \leq 1$.

Eq. 4.4 can then be used to replace the reference to "machine-readable symbolic expressions," i.e. the "MRSE's" in the $v = 3/2$ entry in Part I, Table 4.3 and Part I, Appendix S [1], with the more compact Eq. 4.5, below, where Eq. 4.4's $I_6$, $a$, and $b$ have been identified as $R_{i,j}^{(v=3/2)}$, $x_{i,k}$, and $x_{j,k}$, respectively, as in Part I, Table 4.3.

$$R_{i,j}^{(v=3/2)} = \prod_{k=1}^{d} \frac{1}{24\sqrt{3\theta_k}} \left\{ 2 \begin{bmatrix} 15 & 1 \\ +15|x_{i,k} - x_{j,k}| \sqrt{3\theta_k}^1 \\ + 6|x_{i,k} - x_{j,k}|^2 \sqrt{3\theta_k}^2 \\ + |x_{i,k} - x_{j,k}|^3 \sqrt{3\theta_k}^3 \end{bmatrix} e^{-2\sqrt{3\theta_k} \left| \frac{x_{i,k} - x_{j,k}}{2} \right|} \right. \\ \left. - 3 \mathcal{T}_{x_{i,k}; x_{j,k}}^{(+)} \begin{bmatrix} 5 \\ +3 \begin{pmatrix} 2 \\ +x_{i,k} + x_{j,k} \end{pmatrix} \sqrt{3\theta_k}^1 \\ +2 \begin{pmatrix} 1 \\ +x_{i,k} + x_{j,k} \\ +x_{i,k} x_{j,k} \end{pmatrix} \sqrt{3\theta_k}^2 \end{bmatrix} e^{-2\sqrt{3\theta_k} \left(1 + \frac{x_{i,k} + x_{j,k}}{2}\right)} \right\}. \qquad (4.5)$$

We note the successive integer coefficients in the upper square bracket of Eq. 4.5, viz. $(15, 15, 6, 1)$ are successive Bessel numbers of the first kind [8].

## 5. Integral $I_8$

In similar manner to what was just carried out in Sec. 4, the following is the integral involving the Matérn-covariance function with class parameter 5/2, from Part I, Eq. K.1 [1].

$$I_8 \equiv \frac{1}{2} \int_{-1}^{1} \left[ 1 + \sqrt{5\theta(a-x)^2} + \frac{5\theta(a-x)^2}{3} \right] \left[ 1 + \sqrt{5\theta(b-x)^2} + \frac{5\theta(b-x)^2}{3} \right] e^{-\sqrt{5\theta(a-x)^2} - \sqrt{5\theta(b-x)^2}} \, dx.$$

The symbolic-manipulation-software result was the following, which is truncated, here:

$$I8 := -\frac{1}{1080} \frac{1}{\sqrt{\theta} \, e^{\sqrt{\theta}\sqrt{5}b} \, e^{\sqrt{\theta}\sqrt{5}a} \left(e^{\sqrt{\theta}\sqrt{5}}\right)^2} \left( 810\sqrt{5}\, b\theta + 189\sqrt{5} + 150\sqrt{5}\, b^2 \theta \right.$$

$$- 378 \left(e^{\sqrt{\theta}\sqrt{5}a}\right)^2 \sqrt{5} \left(e^{\sqrt{\theta}\sqrt{5}}\right)^2 + 189 \left(e^{\sqrt{\theta}\sqrt{5}b}\right)^2 \left(e^{\sqrt{\theta}\sqrt{5}a}\right)^2 \sqrt{5}$$

$$+ 1350 \left(e^{\sqrt{\theta}\sqrt{5}a}\right)^2 \left(e^{\sqrt{\theta}\sqrt{5}b}\right)^2 \sqrt{\theta} + 150\sqrt{5}\, \theta^2 + 810\sqrt{5}\, a\theta + 1200\, \theta^{3/2}$$

+ many more rows … .

After typesetting and steps of algebraic reduction, this integral becomes

$$I_8 = -\frac{1}{1080\sqrt{\theta}e^{2\sqrt{5\theta}\left(1+\frac{a+b}{2}\right)}} \left\{ \begin{bmatrix} -378\sqrt{5} \\ +(1890a-1890b)\sqrt{\theta} \\ +(-840\sqrt{5}a^2 - 840\sqrt{5}b^2 + 1680\sqrt{5}ab)\sqrt{\theta}^2 \\ +\begin{pmatrix} 1050a^3 - 1050b^3 - 3150a^2b \\ +3150ab^2 \end{pmatrix}\sqrt{\theta}^3 \\ +\begin{pmatrix} -150\sqrt{5}a^4 - 150\sqrt{5}b^4 + 600\sqrt{5}a^3b \\ -900\sqrt{5}a^2b^2 + 600\sqrt{5}ab^3 \end{pmatrix}\sqrt{\theta}^4 \\ +\begin{pmatrix} -50b^5 + 50a^5 - 250a^4b \\ +500a^3b^2 + 250ab^4 - 500a^2b^3 \end{pmatrix}\sqrt{\theta}^5 \end{bmatrix} e^{2\sqrt{5\theta}(a+1)} \right.$$

$$+ \begin{bmatrix} 189\sqrt{5} \\ +(1350 + 675b + 675a)\sqrt{\theta} \\ +\begin{pmatrix} 150\sqrt{5}a^2 + 810\sqrt{5} + 150\sqrt{5}b^2 \\ +810\sqrt{5}b + 810\sqrt{5}a \\ +510\sqrt{5}ab \end{pmatrix}\sqrt{\theta}^2 \\ +\begin{pmatrix} 600b^2 + 1800b + 600a^2b \\ +600ab^2 + 2400ab + 1200 \\ +600a^2 + 1800a \end{pmatrix}\sqrt{\theta}^3 \\ +\begin{pmatrix} 150\sqrt{5}b^2 + 150\sqrt{5}a^2 + 300\sqrt{5}b \\ +300\sqrt{5}a + 150\sqrt{5} + 600\sqrt{5}ab \\ +150\sqrt{5}a^2b^2 + 300\sqrt{5}a^2b \\ +300\sqrt{5}ab^2 \end{pmatrix}\sqrt{\theta}^4 \end{bmatrix}$$

$$+ \begin{bmatrix} 189\sqrt{5} \\ +(-675a - 675b + 1350)\sqrt{\theta} \\ +\begin{pmatrix} 810\sqrt{5} + 150\sqrt{5}a^2 - 810\sqrt{5}a \\ +150\sqrt{5}b^2 - 810\sqrt{5}b + 510\sqrt{5}ab \end{pmatrix}\sqrt{\theta}^2 \\ +\begin{pmatrix} 1200 + 600a^2 + 600b^2 - 1800a \\ -1800b - 600a^2b \\ -600ab^2 + 2400ab \end{pmatrix}\sqrt{\theta}^3 \\ +\begin{pmatrix} 150\sqrt{5} + 150\sqrt{5}b^2 + 150\sqrt{5}a^2 \\ -300\sqrt{5}a - 300\sqrt{5}b + 150\sqrt{5}a^2b^2 \\ +600\sqrt{5}ab - 300\sqrt{5}ab^2 - 300\sqrt{5}a^2b \end{pmatrix}\sqrt{\theta}^4 \end{bmatrix} e^{2\sqrt{5\theta}(a+b)} \left. \right\}.$$

Multiplying both numerator and denominator of the ultimate equation by $-\sqrt{5}e^{-2\sqrt{5\theta}\left(1+\frac{a+b}{2}\right)}$, remembering $-1 \leq a \leq b \leq 1$, and recognizing $e^{-2\sqrt{5\theta}\left(1+\frac{a+b}{2}\right)} \cdot e^{2\sqrt{5\theta}(a+1)} = e^{-\sqrt{5\theta}|b-a|}$ and $e^{-2\sqrt{5\theta}\left(1+\frac{a+b}{2}\right)} \cdot e^{2\sqrt{5\theta}(a+b)} = e^{-2\sqrt{5\theta}\left(1-\frac{a+b}{2}\right)}$ gives

$$I_8 = \frac{1}{1080\sqrt{5\theta}} \left\{ \begin{bmatrix} 1890 \\ +1890|b-a|\sqrt{5\theta} \\ +840|b-a|^2\sqrt{5\theta}^2 \\ +210|b-a|^3\sqrt{5\theta}^3 \\ +30|b-a|^4\sqrt{5\theta}^4 \\ +2|b-a|^5\sqrt{5\theta}^5 \end{bmatrix} e^{-\sqrt{5\theta}|b-a|} \\ -\mathcal{T}_{a;b}^{(+)} \begin{bmatrix} 945 \\ +675 \begin{pmatrix} 2 \\ +a+b \end{pmatrix} \sqrt{5\theta} \\ +30 \begin{pmatrix} 27 \\ +27a+27b \\ +5a^2+5b^2+17ab \end{pmatrix} \sqrt{5\theta}^2 \\ +120 \begin{pmatrix} 2 \\ +3a+3b \\ +a^2+b^2+4ab \\ +a^2b+ab^2 \end{pmatrix} \sqrt{5\theta}^3 \\ +30 \begin{pmatrix} 1 \\ +2a+2b \\ +a^2+b^2+4ab \\ +2a^2b+2ab^2 \\ +a^2b^2 \end{pmatrix} \sqrt{5\theta}^4 \end{bmatrix} e^{-2\sqrt{5\theta}\left(1+\frac{a+b}{2}\right)} \right\}. \quad (5.1)$$

Eq. 5.1 is invariant under interchange of $a$ and $b$, and thus it is valid not only under the original assumption $b \geq a$ but also for $b < a$. Thus, Eq. 5.1 is valid for arbitrary $a$ and $b$ on the interval $-1 \leq a, b \leq 1$.

Eq. 5.1 can then be used to replace the reference to "machine-readable symbolic expressions," i.e. the "MRSE's" in the $v = 5/2$ entry in Part I, Table 4.3 and Part I, Appendix T [1], with the more compact Eq. 5.2, below, where eq. 5.1's $I_8$, $a$, and $b$ have been identified as $R_{i,j}^{(v=5/2)}$, $x_{i,k}$, and $x_{j,k}$, respectively, as in Part I, Table 4.3.

$$R_{i,j}^{(5/2)} = \prod_{k=1}^{d} \frac{1}{1080\sqrt{5\theta_k}} \left\{ 2 \begin{bmatrix} 945 \\ +945|x_{i,k}-x_{i,j}|\sqrt{5\theta_k} \\ +420|x_{i,k}-x_{i,j}|^2 \sqrt{5\theta_k}^2 \\ +105|x_{i,k}-x_{i,j}|^3 \sqrt{5\theta_k}^3 \\ +15|x_{i,k}-x_{i,j}|^4 \sqrt{5\theta_k}^4 \\ +|x_{i,k}-x_{i,j}|^5 \sqrt{5\theta_k}^5 \end{bmatrix} e^{-2\sqrt{5\theta_k}\left|\frac{x_i-x_j}{2}\right|} \right.$$
$$\left. -15\mathcal{T}_{x_{i,k};x_{j,k}}^{(+)} \begin{bmatrix} 63 \\ +45\begin{pmatrix} 2 \\ +x_{i,k}+x_{j,k} \end{pmatrix} \sqrt{5\theta_k} \\ +2\begin{pmatrix} 27 \\ +27x_{i,k}+27x_{j,k} \\ +5x_{i,k}^2+17x_{i,k}x_{j,k}+5x_{j,k}^2 \end{pmatrix} \sqrt{5\theta_k}^2 \\ +8\begin{pmatrix} 2 \\ +3x_{i,k}+3x_{j,k} \\ +x_{i,k}^2+4x_{i,k}x_{j,k}+x_{j,k}^2 \\ +x_{i,k}^2 x_{j,k}+x_{i,k}x_{j,k}^2 \end{pmatrix} \sqrt{5\theta_k}^3 \\ +2\begin{pmatrix} 1 \\ +2x_{i,k}+2x_{j,k} \\ +x_{i,k}^2+4x_{i,k}x_{j,k}+x_{j,k}^2 \\ +2x_{i,k}^2 x_{j,k}+2x_{i,k}x_{j,k}^2 \\ +x_{i,k}^2 x_{j,k}^2 \end{pmatrix} \sqrt{5\theta_k}^4 \end{bmatrix} e^{-2\sqrt{5\theta_k}\left(1+\frac{x_i+x_j}{2}\right)} \right\}.$$

(5.2)

In similar manner to what we observed in the $\nu = 3/2$ case of Sec. 4, the successive integer coefficients in the upper square bracket of Eq. 5.2, viz. $(945, 945, 420, 105, 15, 1)$, are successive Bessel numbers of the first kind [8].

## 6. The MINOS optimization-software system

MINOS [2-4] is a general-purpose optimizer for constrained optimization problems involving smooth functions. It is designed for large numbers of sparse linear and nonlinear constraints, ideally with objective gradients and constraint gradients provided by the user. Here we have a smooth objective function of a moderate number of variables with just upper and lower bounds as constraints. MINOS uses a quasi-Newton method to find a local minimizer. Objective gradients are approximated by finite differences in this case. For non-convex problems, varying the starting point increases the chance of finding different local minima.

The Appendix provides numerical examples possibly useful for statistical-software development and validation.

## 7. Revision history

v2: "Generalized" was added to the title. Reference 1 was updated. Reference 7 was corrected.

# Appendix. Tabulated IMSPE-Optimal Designs: $[d, n] = [1, 1 \text{ or } 2]$

We now compare IMSPE-optimal designs and their respective IMSPE values for a variety of single-design-domain-variable, i.e. $d = 1$, cases; for a values of the covariance hyperparameter $\theta = 0.1, 1.0$, or 10. The following programs are compared:

*dMINOS+IMSPE:* double-precision, downhill-search software introduced in Sec. 7, with feasibility and tolerance parameters set to $10^{-7}$, and with 15 non-leading-zero digits printed;

*qMINOS+IMSPE*: quad-precision, downhill-search software introduced in Sec. 7, with feasibility and tolerance parameters set to $10^{-15}$, and with 30 non-leading-zero digits printed;

SMS 20: SMS evaluations [6] of a known optimal design, with the digits parameter set to 20, and with 20 non-leading-zero digits printed; and

MS Excel: spreadsheet evaluations [7] of a known optimal design, with 15 non-leading-zero digits printed.

The tables cover the following cases, where $\tilde{p}$ was defined in [1] as $\tilde{p} = 1$ for exponential covariance and $\tilde{p} = 2$ for Gaussian covariance.

Table 1: $[n, \tilde{p}] = [1, 2]$;

Table 2: $[n, \tilde{p}] = [2, 1 \text{ or } 2]$; and

Table 3: $[n, \nu] = [2, 3/2]$.

| d | n | $\tilde{p}$ | θ | Program | Optimal design: [$x_1$] | IMSPE |
|---|---|---|---|---|---|---|
| 1 | 1 | 2 | 10 | dMINOS+IMSPE | [-6.5 × 10$^{-11}$] | 1.43950 52189 8671 |
| | | | | qMINOS+IMSPE | [ 4.2 × 10$^{-18}$] | 1.43950 52189 86714 51872 &93187 1020 |
| | | | | SMS 20 | [ 0.0 ]* | 1.43950 52189 86714 5188 |
| | | | | MS Excel | [ 0.0 ]* | 1.43950 52189 8671 |
| | | | 1 | dMINOS+IMSPE | [ 1.6 × 10$^{-10}$] | 0.50635 17343 75146 |
| | | | | qMINOS+IMSPE | [-1.6 × 10$^{-14}$] | 0.50635 17343 75145 94920 &10651 27736 |
| | | | | SMS 20 | [ 0.0 ]* | 0.50635 17343 75145 94921 |
| | | | | MS Excel | [ 0.0 ]* | 0.50635 17343 75146 |
| | | | 0.1 | dMINOS+IMSPE | [ 1.0 × 10$^{-8}$ ] | 0.06471 33747 28816 4 |
| | | | | qMINOS+IMSPE | [ 1.0 × 10$^{-16}$] | 0.06471 33747 28816 33798 &13834 57779 4 |
| | | | | SMS 20 | [ 0.0 ]* | 0.06471 33747 28816 33800 0 |
| | | | | MS Excel | [ 0.0 ]* | 0.06471 33747 28816 4 |

Table 1. Putatively IMSPE-optimal designs and their respective IMSPE values computed by four programs for three values of the parameter θ and for $[n, \tilde{p}] = [1,2]$. An asterisk denotes that the optimal design is assumed known to be the singleton design point determined via an exercise in Part I [1]. An ampersand in this or subsequent tables denotes line continuations.

| d | n | $\tilde{p}$ | $\theta$ | prec. | Optimal design: $\begin{bmatrix} x_1 \\ x_2 \end{bmatrix}$ | IMSPE |
|---|---|---|---|---|---|---|
| 1 | 2 | 1 | 10 | d | $\begin{bmatrix} -0.42884\ 34788\ 39919\ 52 \\ 0.42884\ 26854\ 92197\ 06 \end{bmatrix}$ | 1.25050 61071 3192 |
| | | | | q | $\begin{bmatrix} -0.42884\ 30765\ 02973\ 73858 \\ \&09133\ 42642\ 83568\ 8 \\ 0.42884\ 30765\ 02926\ 65101 \\ \&99533\ 87262\ 85821\ 1 \end{bmatrix}$ | 1.25050 61071 31920 36875 &72041 2020 |
| | | | 1 | d | $\begin{bmatrix} -0.56261\ 39784\ 48669\ 01 \\ 0.56261\ 29988\ 43545\ 65 \end{bmatrix}$ | 0.35837 23185 81025 |
| | | | | q | $\begin{bmatrix} -0.56261\ 34844\ 80819\ 48598 \\ \&33756\ 53888\ 48723\ 8 \\ 0.56261\ 34844\ 80748\ 86252 \\ \&78743\ 78714\ 52642\ 6 \end{bmatrix}$ | 0.35837 23185 80888 96934 &11193 78167 |
| | | | 0.1 | d | $\begin{bmatrix} -0.59537\ 25936\ 39878\ 16 \\ 0.59537\ 18978\ 81000\ 64 \end{bmatrix}$ | 0.00397 51567 44849 29 |
| | | | | q | $\begin{bmatrix} -0.59537\ 20850\ 98266\ 84621 \\ \&74477\ 37796\ 58910\ 9 \\ 0.59537\ 20850\ 98266\ 70174 \\ \&05818\ 88228\ 89901\ 0 \end{bmatrix}$ | 0.00397 51567 44848 40954 &70612 61536 26 |
| | | 2 | 10 | d | $\begin{bmatrix} -0.45981\ 81626\ 57083\ 75 \\ 0.45981\ 72978\ 55431\ 02 \end{bmatrix}$ | 0.74875 02831 53917 |
| | | | | q | $\begin{bmatrix} -0.45981\ 77205\ 08375\ 26786 \\ \&79290\ 92871\ 67734\ 6 \\ 0.45981\ 77205\ 08375\ 26761 \\ \&62277\ 70131\ 26293\ 9 \end{bmatrix}$ | 0.74875 02831 53859 71998 &29208 74009 |
| | | | 1 | d | $\begin{bmatrix} -0.54798\ 53969\ 00615\ 10 \\ 0.54798\ 43138\ 92301\ 51 \end{bmatrix}$ | 0.10433 80536 94004 |
| | | | | q | $\begin{bmatrix} -0.54798\ 48421\ 86733\ 04008 \\ \&65529\ 12592\ 69386\ 9 \\ 0.54798\ 48421\ 86658\ 24396 \\ \&41341\ 03262\ 85961\ 7 \end{bmatrix}$ | 0.10433 80536 93786 37528 &69587 81117 |
| | | | 0.1 | d | $\begin{bmatrix} -0.57433\ 46652\ 01583\ 16 \\ 0.57433\ 40235\ 86725\ 92 \end{bmatrix}$ | 0.00237 33529 28078 27 |
| | | | | q | $\begin{bmatrix} -0.57433\ 43404\ 66996\ 12822 \\ \&90362\ 32524\ 99364\ 9 \\ 0.57433\ 43404\ 66946\ 06100 \\ \&45162\ 40790\ 45858\ 7 \end{bmatrix}$ | 0.00237 33529 28077 26460 &78477 09326 67 |

Table 2. Putatively IMSPE-optimal designs and their respective IMSPE values computed by *dMINOS+IMSPE* or *qMINOS+IMSPE*, (denoted in the "prec." column as "d" or "q," respectively) for $[n, \tilde{p}] = [2, 1\ \text{or}\ 2]$.

| d | n | $\nu$ | θ | prec. | Optimal design: $\begin{bmatrix}x_1\\x_2\end{bmatrix}$ | IMSPE |
|---|---|---|---|---|---|---|
| 1 | 2 | $\frac{3}{2}$ | 10 | d | $\begin{bmatrix}-0.49931\ 12330\ 51960\ 85\\ 0.49931\ 11448\ 13527\ 29\end{bmatrix}$ | 0.63748 69619 95181 |
| | | | | q | $\begin{bmatrix}-0.49931\ 12231\ 88040\ 38999\\ \&57010\ 32901\ 68733\ 6\\ 0.49931\ 12231\ 88040\ 28693\\ \&87885\ 82708\ 10440\ 2\end{bmatrix}$ | 0.63748 69619 95178 11750 &76602 12266 |
| | | | 1 | d | $\begin{bmatrix}-0.55786\ 61700\ 76245\ 84\\ 0.55786\ 51713\ 08787\ 54\end{bmatrix}$ | 0.12389 32505 77532 |
| | | | | q | $\begin{bmatrix}-0.55786\ 56901\ 81842\ 85558\\ \&43313\ 00198\ 91411\ 8\\ 0.55786\ 56901\ 81842\ 84817\\ \&65631\ 59758\ 32962\ 6\end{bmatrix}$ | 0.12389 32505 77378 38246 &37735 29331 |
| | | | 0.1 | d | $\begin{bmatrix}-0.58014\ 88981\ 52651\ 69\\ 0.58014\ 81621\ 23475\ 97\end{bmatrix}$ | 0.00916 99981 76725 25 |
| | | | | q | $\begin{bmatrix}-0.58014\ 85024\ 91707\ 01493\\ \&66244\ 83632\ 40266\ 2\\ 0.58014\ 85024\ 91659\ 83303\\ \&95527\ 55932\ 86527\ 2\end{bmatrix}$ | 0.00916 99981 76714 41540 &59189 34232 91 |

Table 3. Putatively IMSPE-optimal designs and their respective IMSPE values computed by *dMINOS+IMSPE* or *qMINOS+IMSPE,* (denoted in the "prec." column as "d" or "q," respectively), for $[n, \nu] = [2, 3/2]$.